\documentclass[12pt]{article}
\newcommand{\beq}{\begin{equation}}
\newcommand{\eeq}{\end{equation}}
\newcommand{\beqa}{\begin{eqnarray}}
\newcommand{\eeqa}{\end{eqnarray}}
\newcommand{\beqar}{\begin{eqnarray*}}
\newcommand{\eeqar}{\end{eqnarray*}}

\newcommand{\inn}{\!\cdot\!}

\newcommand{\z}{\zeta}

\newcommand{\eg}{{\it e.g.,}\ }
\newcommand{\ie}{{\it i.e.,}\ }
\newcommand{\labell}[1]{\label{#1}} 
\newcommand{\reef}[1]{(\ref{#1})}
\newcommand\prt{\partial}

\newcommand\tG{{\widetilde G}}

\newcommand\ta{{\tilde a}}
\newcommand\tb{{\tilde b}}
\newcommand\tc{{\tilde c}}
\newcommand\td{{\tilde d}}

\newcommand\ti{{\tilde i}}
\newcommand\tj{{\tilde j}}

\begin{document}

\begin{titlepage}

\begin{center}



\vskip 2 cm
{\LARGE \bf An off-shell   D-brane action at order $ \alpha'^2$\\ \vskip 0.6 cm in flat spacetime  
 }\\
\vskip 1.25 cm
  Mohammad R. Garousi\footnote{garousi@ um.ac.ir}

\vskip 1 cm
{{\it Department of Physics, Ferdowsi University of Mashhad\\}{\it P.O. Box 1436, Mashhad, Iran}\\}
\vskip .1 cm
\vskip .1 cm

\end{center}

\vskip 0.5 cm

\begin{abstract}
\baselineskip=18pt
We use compatibility of  the second fundamental form  corrections to DBI action  at order $\alpha'^2$ which includes trace of the second fundamental form,    with   T-duality and with the linear S-duality   as  guiding principles to find   an off-shell  D-brane action at order $\alpha'^2$  in type II superstring theories in flat spcetime. 
 
\end{abstract}
\end{titlepage}
\section{Introduction and Results}
In flat spacetime with no massless closed string background,     effective action of  a D$_p$-brane in type II superstring theories at long wavelength limit   is given by   Dirac-Born-Infeld (DBI) action \cite{Leigh:1989jq,Bachas:1995kx}
\beqa
S_p&=&-T_p\int d^{p+1}\sigma\sqrt{-\det(\tG_{ab}+  F_{ab})}\labell{DBI}
\eeqa
 where   we normalize the gauge field to  absorb the factor of $2\pi\alpha'$ in front of the gauge field strength $F_{ab}$ which usually appears in the literature. In above equation,  $\tG_{ab}$  is the pull-back of the  bulk flat metric onto the world-volume\footnote{Our index convention is that the Greek letters  $(\mu,\nu,\cdots)$ are  the indices of the space-time coordinates, the Latin letters $(a,d,c,\cdots)$ are the world-volume indices and the letters $(i,j,k,\cdots)$ are the normal bundle indices. The Killing index in the reduction of 10-dimensional spacetime to 9-dimensional spacetime is $y$.}, \ie
\beqa
\tG_{ab}&=&\frac{\prt X^{\mu}}{\prt\sigma^a}\frac{\prt X^{\nu}}{\prt\sigma^b}\eta_{\mu\nu}\nonumber\\
 &=&\eta_{ab}+ \prt_a\chi^{i}\prt_b\chi^{j}\eta_{ij}\labell{pull}
\eeqa
where in the second line the pull-back is in the static gauge.   The DBI action \reef{DBI} is invariant under T-duality and its equations of motion for the case of $p=3$, are invariant under S-duality \cite{Gibbons:1995ap,Tseytlin:1996it,Green:1996qg}.     With our  normalization for the gauge field,  the DBI action is   at the leading order of $\alpha'$.  The first correction  to this action   is at order $\alpha'^2$ in which we are interested in this paper.  The $\alpha'$ corrections to Born-Infeld action  have been studied in \cite{Abouelsaood:1986gd,Tseytlin:1987ww,Andreev:1988cb, Wyllard:2000qe, Andreev:2001xx, Wyllard:2001ye} in the $\sigma$-model approach.

For zero gauge field, the general covariance requires the world-volume couplings at any order of $\alpha'$ consists of various contractions of   the second fundamental form $\Omega$ and its derivatives.    At order $\alpha'^2$, such  couplings have structures $\Omega^4$ or $(D\Omega)^2$. The latter couplings  are not consistent with supersymmetry \cite{Bachas:1999um}, and the former couplings   have been found in \cite{Bachas:1999um,Jalali:2015xca} through the curvature squared couplings to be 
\beqa
S_1&=&-\frac{\pi^2\alpha'^2T_p}{48}\int d^{p+1}\sigma\sqrt{-\det(\tG_{ab}+F_{ab} )}\bigg[  4 \Omega ^a{}^b{}^i \Omega _a{}_b{}^j \Omega ^c{}^d{}_i \Omega _c{}_d{}_j-4 \Omega ^a{}^b{}^i \Omega _a{}^c{}_i \Omega _b{}^d{}^j \Omega _c{}_d{}_j\nonumber\\
&&+4 \Omega ^a{}_a{}^i \Omega ^b{}^c{}_i \Omega _b{}^d{}^j \Omega _c{}_d{}_j -6 \Omega ^a{}_a{}^i \Omega ^b{}_b{}^j \Omega ^c{}^d{}_i \Omega _c{}_d{}_j+2 \Omega ^a{}_a{}^i \Omega ^b{}_b{}_i \Omega ^c{}_c{}^j \Omega ^d{}_d{}_j \bigg]\labell{DBI1} 
\eeqa
The second fundamental form   in flat spacetime  reduces to    the acceleration (see \eg  \cite{Corley:2001hg}), \ie $\Omega_{ab}{}^i=\prt_a\prt_b\chi^i$.  The world-volume indices in above action are raised by the inverse of the pull-back metric, $\tG^{ab}$, and the transverse indices are lowered by  
\beqa
\widetilde{\bot}_{ij}&=&\eta_{ij}-\eta_{ik}\eta_{jl}\prt_a\chi^k\prt_b\chi^l\tG^{ab}
\eeqa
 The   couplings \reef{DBI1} are consistent with the S-matrix element of four transverse scalar fields at order $\alpha'^2$ \cite{Bachas:1999um}, however, they include the trace of the second fundamental form which is zero on-shell. They have been found in \cite{Jalali:2015xca} by studying the consistency of the couplings of one massless closed and two open strings under T-duality and linear S-duality transformations. The action \reef{DBI1} contains   couplings at all orders of the scalar fields. 

It is known that the S-matrix elements of four massless NS states have no massless pole at order $\alpha'^2$. On the other hand, the S-matrix elements satisfy the Ward identity corresponding to the S-duality and T-duality \cite{Garousi:2011we}, \ie the amplitudes are invariant under   linear dualities. The T-duality transformations for the massless NS states are linear whereas the S-duality transformation for the gauge field   is nonlinear \cite{Gibbons:1995ap}. One may expect then the four off-shell  NS couplings at order $\alpha'^2$ to be invariant under   T-duality and/or   linear S-duality. We are going to impose these constraints to include four non-constant  gauge field strengths into the action \reef{DBI1}. The T-duality constraint on higher derivative couplings of branes have been studied in \cite{Garousi:2009dj,Becker:2010ij,Godazgar:2013bja}. Four-field on-shell couplings to all orders in $\alpha'$ has been obtained in \cite{deRoo:2003xv} from the corresponding S-matrix element.

Four-derivative on-shell corrections to the DBI action involving four gauge field strengths
and four derivatives have been known for a long time \cite{Andreev:1988cb}. They do not include the couplings which are related by T-duality  to the off-shell couplings in \reef{DBI1}.  To include these terms, one may constrain the couplings to be invariant under  the dualities,   reduces to \reef{DBI1} when the gauge fields are zero, and to be consistent with the corresponding S-matrix element in string theory. There are two ways to impose the dualities. One may first consider all  S-duality invariant couplings with unknown coefficients and then imposes the T-duality constraint and the reduction to \reef{DBI1}  to find the coefficients. Or, one may consider all arbitrary couplings with unknown coefficients and imposes the T-duality constraint, the reduction to \reef{DBI1} and then  the S-duality constraint. It turns out that the two results are not identical.  In fact, it turns out that in the latter approach the S-duality can be imposed only on on-shell couplings. In other words, the T-duality and the linear S-duality do not commute. However,  the couplings in both cases reproduce the same S-matrix element, so they are    identical up to a field redefinition.

In the first approach, one has to consider all linear S-duality invariant couplings. For two gauge fields, one may  consider the   tensor $Q_{abcdef}$  which is defined by the following  expression:
\beqa
Q_{abcdef}&\equiv&\prt_a F_{bc}\prt_d F_{ef}+\prt_a (*F)_{bc}\prt_d (*F)_{ef}\labell{Q}
\eeqa
where  $(*F)_{ab}=\epsilon_{abcd}F^{cd}/2$.  It it invariant under the linear S-duality for D$_3$-brane. To extend the couplings involving this tensor to the   arbitrary D$_p$-brane, one has to replace the contraction of two four-dimensional Levi-Civita tensors in terms of metric to produce the corresponding couplings on the world-volume of D$_3$-brane, and then extend the couplings to the arbitrary dimensions.
For four gauge fields, however, the S-duality invariant expression which includes two $Q$'s,  involves  four   Levi-Civita tensors. There are three different parings of these tensors. In   section 3, we will show that the different parings   produce different expressions for contractions of four gauge fields. Apart from this ambiguity, one can consider all contractions of $Q$ and $\Omega$ at order $\alpha'^2$ with unknown coefficients, and impose the T-duality constraint and the reduction to \reef{DBI1} to find the coefficients. For the specific paring of the  Levi-Civita tensors that the two  Levi-Civita tensors in each $Q$ contract with each other, we have done the calculation and found that above constraints fix the constants such that the couplings satisfy the S-matrix element without any further constraints on the coefficients. This approach can not be extended to more than four-field   couplings because the contact terms of the corresponding S-matrix elements do not satisfy the S-dual Ward identity (see   section 3).

	In the second approach, one has to consider all contractions of $\Omega$ and $\prt F$  at order $\alpha'^2$ with unknown coefficients and constrain them to be consistent with the T-duality and with the four-field couplings in \reef{DBI1}. Imposing these constraints, one finds that the resulting off-shell couplings do not satisfy the S-duality constraint even at two gauge field level. So we are forced to impose   on-shell S-duality. On the other hand, the 4-point S-matrix element satisfy the S-dual Ward identity \cite{Garousi:2011vs}, so to impose the on-shell S-duality, one may impose the consistency of the couplings with S-matrix element. This gives one extra constraint.   Up to some total derivative terms and the Bianchi identity, we have found    the following couplings between two scalar fields and two gauge fields:
\beqa
S_2&\!\!\!\!\!=\!\!\!\!\!&-\frac{\pi^2\alpha'^2T_p}{48}\int d^{p+1}\sigma\sqrt{-\det(\tG_{ab}+F_{ab} )}\bigg[ 
 4 \Omega_c {} _d {}^i \Omega^
   e {} _e {} _i \prt_aF _b {}^d \prt^aF^b {}^c -8 \Omega_c {}^e {}^
   i \Omega_d {} _e {} _i \prt_aF _b {}^d \prt^aF^b {}^c\nonumber\\&&+ 
 12 \Omega_c {}^e {}^
   i \Omega_d {} _e {} _i \prt^aF^b {}^c \prt_bF _a {}^d - 
 4 \Omega_c {} _d {}^i \Omega^
   e {} _e {} _i \prt^aF^b {}^c \prt_bF _a {}^d - 
 4 \Omega^d {} _d {}^i \Omega^
   e {} _e {} _i \prt^aF_a {}^b \prt^cF_b {} _c \nonumber\\&&- 
 4 \Omega_c {}^e {}^
   i \Omega_d {} _e {} _i \prt^aF_a {}^b \prt^cF_b {}^d + 
 12 \Omega_c {} _d {}^i \Omega^
   e {} _e {} _i \prt^aF_a {}^b \prt^cF_b {}^d + 
 8 \Omega_a {} _c {}^
   i \Omega_d {} _e {} _i \prt^aF^b {}^c \prt^dF_b {}^e \bigg]\labell{DBI2} 
\eeqa
  The world-volume indices in \reef{DBI2}  are raised by  $\eta^{ab}$, and the transverse indices are lowered by $\eta_{ij}$. 
	 The above   constraints also fix  the following  couplings between four gauge fields:
\beqa
S_3&\!\!\!\!\!=\!\!\!\!\!&-\frac{\pi^2\alpha'^2T_p}{48}\int d^{p+1}\sigma\sqrt{-\det(\tG_{ab} +F_{ab})}\bigg[ 2 \prt^aF^b {}^c \prt_bF _a {}^d \prt_cF^e {}^f \prt_dF _e {} _f  \labell{DBI3}\\&&- 
 \frac{1}{2} \prt_aF _b {} _c \prt^aF^b {}^c \prt_dF _e {} _f \prt^dF^e {}^
   f - 2 \prt^aF_a {}^b \prt^cF_b {} _c \prt_dF _e {} _f \prt^dF^e {}^
   f - 4 \prt^aF^b {}^c \prt_bF _c {}^d \prt_dF^e {}^
    f \prt_eF _a {} _f\nonumber\\&& + 
 2 \prt_aF^d {}^e \prt^aF^b {}^c \prt_fF _c {} _e \prt^f F_b {} _d + 
 2 \prt^aF_a {}^b \prt^cF_b {}^d \prt_dF _c {}^e \prt^f F_e {} _f + 
 6 \prt^aF_a {}^b \prt_cF^e {}^f \prt^cF_b {}^d \prt_dF _e {} _f\bigg]\nonumber
\eeqa
The couplings \reef{DBI2} and  \reef{DBI3} are the T-dual completion of four scalar couplings in \reef{DBI1} and are consistent with on-shell linear S-duality. The above actions are consistent with the S-matrix elements of four massless NS vertex operators, however, there are many couplings in these actions that are zero on-shell. They have been found   by the duality constraints.  Unlike the first approach, this approach can be extended to higher order fields. It would be interesting then to find the T-dual completion of all infinite scalar fields in \reef{DBI1}.   

\section{Calculations}

In this section, using the Mathematica package ``xAct'' \cite{CS}, we are going to write all four-field couplings of gauge field and/or transverses scalar fields    with unknown coefficients. We then constrain the coefficients by imposing the consistency of the couplings with the  dualities. Since the known scalar couplings \reef{DBI1} involve only second derivative of the scalar fields and T-duality transforms the gauge field along the Killing direction to the scalar field in the dual theory, \ie $A_y\rightarrow\chi^y$, we expect the four gauge field couplings to have structure $(\prt F)^4$ and two gauge field   two scalar couplings to have structure $(\prt F)^2\Omega^2$. In the first approach, one first uses the S-duality constraint and then the  T-duality constraint, whereas in the second approach one first uses the T-duality and then the on-shell S-duality constraint.  The calculations in both cases are similar so in this section we  illustrate how we got the results in \reef{DBI2} and \reef{DBI3} in the second approach. In section 3, we discuss the couplings in the first approach.

So in the second approach, we have to consider all contractions of $\Omega$ and $\psi_{abc}\equiv\prt_a F_{bc}$ at order $\alpha'^2$. Using  ``xAct'' , one finds the following  65 different contractions:  
\beqa
&\!\!\!\!\!\!\!\!\!\!\!\!\!\!&S=-\frac{\pi^2T_p\alpha'^2}{48}\int d^{p+1}\sigma\sqrt{-\det(\tG_{ab}+F_{ab} )}\bigg[   c_{1} \psi_a {}^d {}^e \psi^a {}^b {}^c \psi_b {} _d {}^f \psi_c {} _e {} _f + 
 c_{2} \psi_a {}^d {}^e \psi^a {}^b {}^c \psi_b {} _c {}^f \psi_d {} _e {} _f\nonumber\\&\!\!\!\!\!\!\!\!\!\!\!\!\!\!& + 
 c_{3} \psi_a {} _b {}^d \psi^a {}^b {}^c \psi_c {}^e {}^f \psi_d {} _e {} _f + 
 c_{4} \psi^a {}^b {}^c \psi_b {} _a {}^d \psi_c {}^e {}^f \psi_d {} _e {} _f + 
 c_{5} \psi^a {} _a {}^b \psi_c {}^e {}^f \psi^c {} _b {}^d \psi_d {} _e {} _f + 
 c_{6} \psi_a {} _b {} _c \psi^a {}^b {}^c \psi_d {} _e {} _f \psi^d {}^e {}^f\nonumber\\&\!\!\!\!\!\!\!\!\!\!\!\!\!\!&  + 
 c_{7} \psi^a {} _a {}^b \psi^c {} _b {} _c \psi_d {} _e {} _f \psi^d {}^e {}^f +
  c_{8} \psi^a {}^b {}^c \psi_b {}^d {}^e \psi_c {} _d {}^f \psi_e {} _a {} _f + 
 c_{9} \psi^a {}^b {}^c \psi_b {} _c {}^d \psi_d {}^e {}^f \psi_e {} _a {} _f + 
 c_{10} \psi_a {}^d {}^e \psi^a {}^b {}^c \psi_b {} _d {}^f \psi_e {} _c {} _f \nonumber\\&\!\!\!\!\!\!\!\!\!\!\!\!\!\!& + 
 c_{11} \psi^a {} _a {}^b \psi^c {} _b {}^d \psi_d {}^e {}^f \psi_e {} _c {} _f +
  c_{12} \psi_a {} _b {}^d \psi^a {}^b {}^c \psi_c {}^e {}^f \psi_e {} _d {} _f +
  c_{13} \psi^a {}^b {}^c \psi_b {} _a {}^d \psi_c {}^e {}^f \psi_e {} _d {} _f +
  c_{14} \psi^a {} _a {}^b \psi_b {}^c {}^d \psi_c {}^e {}^f \psi_e {} _d {} _f\nonumber\\&\!\!\!\!\!\!\!\!\!\!\!\!\!\!&  +
  c_{15} \psi^a {} _a {}^b \psi_c {}^e {}^f \psi^c {} _b {}^
   d \psi_e {} _d {} _f + 
 c_{16} \psi_a {} _b {} _c \psi^a {}^b {}^c \psi^d {}^e {}^f \psi_e {} _d {} _f +
  c_{17} \psi^a {}^b {}^c \psi_b {} _a {} _c \psi^d {}^e {}^f \psi_e {} _d {} _f + 
 c_{18 }\psi^a {} _a {}^b \psi^c {} _b {} _c \psi^d {}^e {}^
   f \psi_e {} _d {} _f \nonumber\\&\!\!\!\!\!\!\!\!\!\!\!\!\!\!& + 
 c_{19} \psi_a {} _b {}^d \psi^a {}^b {}^c \psi_e {} _d {} _f \psi^e {} _c {}^f +
  c_{20} \psi^a {} _a {}^b \psi^c {} _b {}^d \psi_e {} _d {} _f \psi^e {} _c {}^
   f + c_{21} \psi^a {}^b {}^c \psi_b {}^d {}^e \psi_d {} _c {}^
   f \psi_f {} _a {} _e + 
 c_{22} \psi_a {}^d {}^e \psi^a {}^b {}^c \psi_b {} _d {}^f \psi_f {} _c {} _e \nonumber\\&\!\!\!\!\!\!\!\!\!\!\!\!\!\!& + 
 c_{23} \psi_a {}^d {}^e \psi^a {}^b {}^c \psi_b {} _c {}^f \psi_f {} _d {} _e + 
 c_{24} \psi_a {} _b {}^d \psi^a {}^b {}^c \psi^e {} _c {}^f \psi_f {} _d {} _e +
  c_{25} \psi^a {}^b {}^c \psi_b {} _a {}^d \psi^e {} _c {}^
   f \psi_f {} _d {} _e + 
 c_{26} \psi^a {} _a {}^b \psi^c {} _b {}^d \psi^e {} _c {}^
   f \psi_f {} _d {} _e \nonumber\\&\!\!\!\!\!\!\!\!\!\!\!\!\!\!& + 
 c_{27} \psi_a {}^d {}^e \psi^a {}^b {}^c \psi_f {} _d {} _e \psi^f {} _b {} _c +
  c_{28} \psi_a {}^d {}^e \psi^a {}^b {}^c \psi_f {} _c {} _e \psi^
   f {} _b {} _d + 
 c_{29} \psi^a {} _a {}^b \psi_b {}^c {}^d \psi_c {} _d {}^e \psi^f {} _e {} _f +
  c_{30} \psi^a {} _a {}^b \psi_c {} _d {}^e \psi^c {} _b {}^d \psi^
   f {} _e {} _f \nonumber\\&\!\!\!\!\!\!\!\!\!\!\!\!\!\!& + 
 c_{31} \psi^a {} _a {}^b \psi^c {} _b {}^d \psi_d {} _c {}^e \psi^
   f {} _e {} _f + 
 c_{32} \psi^a {} _a {}^b \psi^c {} _b {} _c \psi^d {} _d {}^e \psi^
   f {} _e {} _f + 
 c_{33} \psi^a {} _a {}^b \psi_b {}^c {}^d \psi^e {} _c {} _d \psi^
   f {} _e {} _f + 
 c_{34} \psi^a {}^b {}^c \psi^d {} _b {}^e \Omega_a {} _e {}^
   i \Omega_c {} _d {} _i \nonumber\\&\!\!\!\!\!\!\!\!\!\!\!\!\!\!& + 
 c_{35} \Omega_a {}^c {}^j \Omega^a {}^b {}^
   i \Omega_b {}^d {} _j \Omega_c {} _d {} _i + 
 c_{36} \Omega_a {}^c {} _i \Omega^a {}^b {}^
   i \Omega_b {}^d {}^j \Omega_c {} _d {} _j + 
 c_{37} \Omega^a {} _a {}^i \Omega_b {}^d {}^
   j \Omega^b {}^c {} _i \Omega_c {} _d {} _j + 
 c_{38} \psi^a {}^b {}^c \psi_b {}^d {}^e \Omega_a {} _d {}^
   i \Omega_c {} _e {} _i \nonumber\\&\!\!\!\!\!\!\!\!\!\!\!\!\!\!& + 
 c_{39} \psi^a {}^b {}^c \psi^d {} _b {}^e \Omega_a {} _d {}^
   i \Omega_c {} _e {} _i + 
 c_{40} \psi_a {}^d {}^e \psi^a {}^b {}^c \Omega_b {} _d {}^
   i \Omega_c {} _e {} _i + 
 c_{41} \psi^a {} _a {}^b \psi^c {}^d {}^e \Omega_b {} _d {}^
   i \Omega_c {} _e {} _i + 
 c_{42} \Omega_a {} _b {}^j \Omega^a {}^b {}^
   i \Omega_c {} _d {} _j \Omega^c {}^d {} _i \nonumber\\&\!\!\!\!\!\!\!\!\!\!\!\!\!\!& + 
 c_{43} \Omega^a {} _a {}^i \Omega^b {} _b {}^
   j \Omega_c {} _d {} _j \Omega^c {}^d {} _i + 
 c_{44} \Omega_a {} _b {} _i \Omega^a {}^b {}^
   i \Omega_c {} _d {} _j \Omega^c {}^d {}^j + 
 c_{45} \Omega^a {} _a {}^i \Omega^
   b {} _b {} _i \Omega_c {} _d {} _j \Omega^c {}^
   d {}^j + 
 c_{46} \psi^a {}^b {}^c \psi^d {} _b {}^e \Omega_a {} _c {}^
   i \Omega_d {} _e {} _i \nonumber\\&\!\!\!\!\!\!\!\!\!\!\!\!\!\!& + 
 c_{47} \psi^a {}^b {}^c \psi_b {} _c {}^d \Omega_a {}^e {}^
   i \Omega_d {} _e {} _i + 
 c_{48} \psi^a {}^b {}^c \psi^d {} _b {} _c \Omega_a {}^e {}^
   i \Omega_d {} _e {} _i + 
 c_{49} \psi^a {} _a {}^b \psi^c {} _c {}^d \Omega_b {}^e {}^
   i \Omega_d {} _e {} _i + 
 c_{50 }\psi_a {} _b {}^d \psi^a {}^b {}^c \Omega_c {}^e {}^
   i \Omega_d {} _e {} _i \nonumber\\&\!\!\!\!\!\!\!\!\!\!\!\!\!\!& + 
 c_{51} \psi^a {}^b {}^c \psi_b {} _a {}^d \Omega_c {}^e {}^
   i \Omega_d {} _e {} _i + 
 c_{52} \psi^a {} _a {}^b \psi^c {} _b {}^d \Omega_c {}^e {}^
   i \Omega_d {} _e {} _i + 
 c_{53} \Omega^a {} _a {}^i \Omega^
   b {} _b {} _i \Omega^c {} _c {}^j \Omega^
   d {} _d {} _j + 
 c_{54} \psi_a {} _b {} _c \psi^a {}^b {}^
   c \Omega_d {} _e {} _i \Omega^d {}^e {}^i \nonumber\\&\!\!\!\!\!\!\!\!\!\!\!\!\!\!& + 
 c_{55} \psi^a {}^b {}^
   c \psi_b {} _a {} _c \Omega_d {} _e {} _i \Omega^
   d {}^e {}^i + 
 c_{56} \psi^a {} _a {}^b \psi^
   c {} _b {} _c \Omega_d {} _e {} _i \Omega^d {}^e {}^i + 
 c_{57} \psi^a {}^b {}^c \psi_b {} _c {}^d \Omega_a {} _d {}^
   i \Omega^e {} _e {} _i + 
 c_{58} \psi^a {}^b {}^c \psi^d {} _b {} _c \Omega_a {} _d {}^
   i \Omega^e {} _e {} _i \nonumber\\&\!\!\!\!\!\!\!\!\!\!\!\!\!\!& + 
 c_{59} \psi^a {} _a {}^b \psi^c {} _c {}^d \Omega_b {} _d {}^
   i \Omega^e {} _e {} _i + 
 c_{60} \psi_a {} _b {}^d \psi^a {}^b {}^c \Omega_c {} _d {}^
   i \Omega^e {} _e {} _i + 
 c_{61} \psi^a {}^b {}^c \psi_b {} _a {}^d \Omega_c {} _d {}^
   i \Omega^e {} _e {} _i + 
 c_{62} \psi^a {} _a {}^b \psi^c {} _b {}^d \Omega_c {} _d {}^
   i \Omega^e {} _e {} _i\nonumber\\&\!\!\!\!\!\!\!\!\!\!\!\!\!\!&  + 
 c_{63} \psi_a {} _b {} _c \psi^a {}^b {}^c \Omega^d {} _d {}^
   i \Omega^e {} _e {} _i + 
 c_{64} \psi^a {}^b {}^c \psi_b {} _a {} _c \Omega^d {} _d {}^
   i \Omega^e {} _e {} _i + 
 c_{65} \psi^a {} _a {}^b \psi^c {} _b {} _c \Omega^d {} _d {}^
   i \Omega^e {} _e {} _i \bigg]\labell{DBI4}
\eeqa 
where $c_1,\cdots c_{65}$ are unknown coefficients. In writing the above couplings we have used the mono-term symmetries of $\Omega$ and $\psi$, \ie  the second fundamental form   is symmetric with respect to its first two indices and   $\psi$ is antisymmetric with respect to last two indices. 
The tensor $\psi$ has also   multi-term symmetry, \ie the Bianchi identity, which is not imposed in \reef{DBI4}. Moreover, for a specific relation between the coefficients, some combinations of the above couplings are total derivative terms.
Using the Bianchi identity and ignoring some total derivative terms, one can find some relations between the coefficients  in \reef{DBI4}. One may try to find such relations and   use them to reduce the couplings in \reef{DBI4} to independent ones. Then imposes the T-duality constraint to find the relations between the coefficients of the independent terms. Alternatively, one may first use the T-duality constraint to find   relations between the coefficients in \reef{DBI4} and then impose   the Bianchi identity and remove the total derivative terms. In this paper, we follow the latter approach which is much easier to do with computer, as we will see shortly.  

To constraint the above couplings to be consistent with T-duality, following \cite{Garousi:2009dj}, we   reduce the 10-dimensional space-time to the 9-dimensional space-time. It reduces \reef{DBI4}  to  two  different actions $S_p^w$ and  $S_p^t$. In $S_p^w$,  the Killing direction $y$ is a world-volume direction, \ie $a=(\tilde{a},y)$  and in $S_p^t$   the Killing direction   is a transverse direction, $i=(\tilde{i},y)$. The transformation of  $S_p^w$ under the   T-duality   which is called $S_{p-1}^{wT}$, may be equal to $S_{p-1}^t$ up to some total derivative terms which must be ignored in the action, \ie
\beqa
S_{p-1}^{wT}-S_{p-1}^t&=&0\labell{Tconstraint}
\eeqa
This constrains the unknown coefficients in the  Lagrangian \reef{DBI4}. Note that if one does not ignore the total derivative terms, then one would find some unnecessary   constraints which make the total derivative terms in the $p$-dimensions to be T-duality invariant. To drop the total derivative terms, we transform the terms in $S_{p-1}^{wT}-S_{p-1}^t$ to the momentum space. This labels fields and their momenta. For the identical fields we have to symmetrize the labels as well, \ie $\prt^a A^b\prt^c A^d B^eC^f$ transforms to 
\beqa
\int d^{p}p_1d^pp_2d^pp_3d^pp_4e^{i p_1\cdot x+i p_2\cdot x+i p_3\cdot x+i p_4\cdot x}\frac{1}{2}(-p_1^aA_1^bp_2^cA_2^d-p_2^aA_2^bp_1^cA_1^d)B_3^eC_4^f
\eeqa
 The integral $\int d^px$ in the action, then produces a delta function imposing the conservation of momentum $\delta^p(p_1+p_2+p_3+p_4)$.
Using this, one then write $p_4$ in terms of $-p_1-p_2-p_3$. This step drops all terms in   $S_{p-1}^{wT}-S_{p-1}^t$ that are total derivatives.

We have found that the above T-duality constraint   gives the following 21 equations between the constants:
\beqa
 &&c_{41}= -c_{20}-c_{26}+c_{38}+c_{39}+2 c_{40},c_{42}= -c_{10}+c_{19}-c_{21}-c_{22}+c_{24}+c_{25}+2 c_{28}-c_{35},\nonumber\\&&c_{43}= 2 c_{10}+2 c_{21}+2 c_{22}-4 c_{28}+c_{30}+c_{31}+2 c_{35},c_{46}= -2 c_{10}+2 c_{19}-2 c_{21}-2 c_{22}\nonumber\\&&+2 c_{24}+2 c_{25}+4 c_{28}-c_{34}-c_{39},c_{49}= c_{29}+c_{30}-2 c_{33}-\frac{c_{38}}{2}-\frac{c_{39}}{2}-c_{40},\nonumber\\&&c_{5}= -2 c_{10}-\frac{c_{11}}{2}-\frac{c_{15}}{2}-\frac{c_{20}}{2}-2 c_{21}-2 c_{22}+4 c_{28}-2 c_{35}-\frac{c_{37}}{2},\nonumber\\&&c_{50}= 2 c_{34}-4 c_{35}+2 c_{36}+2 c_{39}+c_{47}-2 c_{48},c_{51}= c_{12}+c_{13}+2 c_{19}+c_{24}+2 c_{3}-2 c_{34}\nonumber\\&&+4 c_{35}-2 c_{36}+\frac{c_{38}}{2}-\frac{3 c_{39}}{2}+2 c_{4}+c_{40}-c_{47}+2 c_{48},c_{52}= 2 c_{34}-4 c_{35}-c_{37}+2 c_{39},\nonumber\\&&c_{53}= -\frac{c_{10}}{2}-\frac{c_{21}}{2}-\frac{c_{22}}{2}+c_{28}+c_{32}-\frac{c_{35}}{2},c_{55}= -\frac{c_{34}}{2}+c_{35}-\frac{c_{39}}{2}+2 c_{44}-2 c_{54},\nonumber\\&&c_{56}= -\frac{c_{34}}{2}+c_{35}-\frac{c_{39}}{2}-c_{45},c_{59}= \frac{c_{38}}{2}+\frac{c_{39}}{2}+c_{40},c_{6}= \frac{c_{10}}{8}-\frac{c_{16}}{2}-\frac{c_{17}}{4}+\frac{c_{21}}{8}+\frac{c_{22}}{8}\nonumber\\&&-\frac{c_{28}}{4}+\frac{c_{35}}{8}+\frac{c_{44}}{4},c_{60}= -2 c_{34}+4 c_{35}+c_{37}-2 c_{39}+c_{57}-2 c_{58},c_{61}= -c_{20}-c_{26}+2 c_{34}\nonumber\\&&-4 c_{35}-c_{37}-\frac{c_{38}}{2}+\frac{3 c_{39}}{2}-c_{40}-c_{57}+2 c_{58},c_{62}= -4 c_{10}-4 c_{21}-4 c_{22}+8 c_{28}-2 c_{30}-2 c_{31}\nonumber\\&&-2 c_{34}-2 c_{39},c_{64}= \frac{c_{34}}{2}-c_{35}+\frac{c_{39}}{2}+c_{45}-2 c_{63},c_{65}= c_{10}+c_{21}+c_{22}-2 c_{28}-2 c_{32}+\frac{c_{34}}{2}\nonumber\\&&+\frac{c_{39}}{2},c_{7}= \frac{c_{10}}{2}-\frac{c_{18}}{2}+\frac{c_{21}}{2}+\frac{c_{22}}{2}-c_{28}+\frac{c_{35}}{2}-\frac{c_{45}}{2},c_{9}= 2 c_{10}+c_{12}+c_{19}+c_{2}+2 c_{21}\nonumber\\&&+2 c_{22}-2 c_{23}+4 c_{27}-4 c_{28}+2 c_{3}+2 c_{35}-c_{36}  \labell{cons2}
\eeqa
Replacing the above constraints into the action \reef{DBI4}, one finds 44  T-dual multiplets.  

These 44 multiplets must be reduced to the four scalar couplings in \reef{DBI1}.   This produces the following 8 equations:
\beqa
&&c_{28}= \frac{c_{10}}{2}-\frac{c_{19}}{2}+\frac{c_{21}}{2}+\frac{c_{22}}{2}-\frac{c_{24}}{2}-\frac{c_{25}}{2}+2,c_{31}= -2 c_{19}-2 c_{24}-2 c_{25}-c_{30}+2,\nonumber\\&&c_{32}= \frac{c_{19}}{2}+\frac{c_{24}}{2}+\frac{c_{25}}{2},c_{35}= 0,c_{36}= -4,c_{37}= 4,c_{44}= 0,c_{45}= 0  \labell{cons3}
\eeqa
Replacing these constraints into the 44 multiplets, one finds 36 T-dual multiplets which reduce to the four scalar couplings in \reef{DBI1}. 

Now we have to impose the consistency with the linear S-duality. There is no ambiguity in imposing the S-duality constraint on two gauge fields, \ie one has to replace $F$ with $*F$ and then replace the two Levi-Civita tensors in terms of metric. The result must be identical to the original couplings. We have found this produces constraints which are not consistent with the constraints in \reef{cons2} and \reef{cons3}. That means the S-duality of equations of motion can not be prompted to the action level in this case. So we impose the linear S-duality on on-shell couplings. On the other hand, the on-shell couplings must be identical to the contact terms of the corresponding S-matrix element. It has been shown in \cite{Garousi:2011vs} that the S-matrix element of four NS vertex operators satisfy linear T-duality. So to impose the on-shell linear S-duality, one may impose the consistency of the above 36 T-dual multiplets with the S-matrix element. This produces the following constraint:  (see Appendix) 
\beqa
c_{4}= -\frac{c_{12}}{2}-\frac{c_{13}}{2}-c_{19}-\frac{c_{24}}{2}-c_{3}+2
\eeqa
Replacing this constraint into the above 36 multiplets, one finds one multiplet with specific coefficients, \ie the couplings in \reef{DBI1}, \reef{DBI2} and \reef{DBI3},  and 35 multiplets with unknown overall coefficients.

 Now we are going to show that, after using  the Bianchi identity and dropping total derivative terms, all the 35 multiplets with unknown coefficients are reduced to 2  unphysical multiplets.  To impose the Bianchi identity, we write the gauge field strength $\prt_aF_{bc}$  in terms of the gauge field, \ie $\prt_a\prt_bA_c-\prt_a\prt_cA_b$. This makes     23 multiplets with the   coefficients 
$c_{1},c_{2},c_{3},c_{12},c_{13},c_{11},c_{16},c_{17},c_{18},c_{8},c_{27},c_{10},c_{14},c_{15},c_{22},c_{48},c_{47},c_{54},c_{57},c_{58},c_{63},c_{21},c_{23}$
to be zero. That means, these constants are the coefficients of terms that are zero by the Bianchi identity.    Therefore, it is safe to set these 23 coefficients to zero. Moreover, after using the Bianchi identity, the terms with coefficients $c_{26} , c_{20}, c_{29}, c_{30}, c_{33}$ appears only as $(c_{20}+c_{26})$ and $(c_{29}+c_{30}-2c_{33})$. That means there are only two independent coefficients. So we set $c_{26}, c_{30},c_{33}$ to zero.

The terms with coefficients $c_{39},c_{40},c_{38},c_{25},c_{19},c_{34},c_{24}$ are total derivative terms. So these 7 coefficients can also be set to zero. Therefore, there are  only 2 independent extra multiplets which are

\beqa
&&-c_{20} \Omega_c {} _d {}^i \Omega^
   e {} _e {} _i \prt^a F^b {}^c \prt_b F_a {}^d + 
 c_{29} \Omega_b {}^e {}^
   i \Omega_d {} _e {} _i \prt^a F_a {}^b \prt^c F_c {}^d - 
 c_{20} \Omega_b {} _d {}^
   i \Omega_c {} _e {} _i \prt^a F_a {}^b \prt^c F^d {}^e \labell{c14}\\&&- 
 \frac{1}{2} c_{20} \prt^a F_a {}^b \prt_c F^e {}^f \prt^c F_b {}^d \prt_d
    F_e {} _f + 
 c_{20} \prt^a F_a {}^b \prt^c F_b {}^d \prt_e F_d {} _f \prt^e F_c {}^
   f + c_{29} \prt^a F_a {}^b \prt_b F^c {}^d \prt_c F_d {}^e \prt^
    f F_e {} _f\nonumber
\eeqa
As can be seen, these multiplets   have no  couplings of four transverse scalar fields, so they are not related to the couplings in \reef{DBI1}.  It is also obvious that the 4-point S-matrix element of above  couplings  are zero, because they involve either the trace of the second fundamental form or $\prt^a F_a {}^b$  which are zero on-shell.

We speculate that the above  multiplets do not   produce any  other S-matrix element. To give one more example, we calculate the S-matrix element of two scalars, one gauge field and one B-field. This amplitude is given by the following Feynman rule:
\beqa
\cal{A}&=&V(B_4,A)G(A)V(A,A_3,\chi_2,\chi_1)+V(B_4,A_3,\chi_2,\chi_1)
\eeqa
where $\chi_1,\chi_2, A_3,B_4$ are the polarizations of the external fields. The symmetry of string theory under the gauge transformation  $A_a\rightarrow A_a-\Lambda_a$   requires one to extend the gauge field strength in effective actions  to $F+B$. Using this replacement in \reef{c14}, one can calculate the vertices $V(B_4,A_3,\chi_2,\chi_1)$ and $V(A,A_3,\chi_2,\chi_1)$, and the vertex $V(B_4,A)$ and propagator $G(A)$ can be calculated from the DBI action. We have done this calculations in details and found zero result.  We expect similar result for any other S-matrix element. Therefore, the multiplets \reef{c14} are   unphysical multiplets. This ends our illustration that all 35 T-dual multiplets are either unphysical or they become total derivatives  after using the Bianchi identity.  

\section{Discussion}

In this paper, we have shown that the consistency of all  four massless NS couplings at order $\alpha'^2$ with T-duality, on-shell linear S-duality and with the scalar couplings in \reef{DBI1}    fixes all couplings up to two unknown coefficients.  The multiplet with the known coefficient   are given in \reef{DBI2} and \reef{DBI3}. These couplings include many terms which  are zero on-shell. They may be removed by using appropriate field redefinitions. The field redefinition, however, change the standard form of the T-duality  transformations. In the specific field variables in which the T-duality transformation is $A_y\rightarrow \chi^y$, the off-shell terms must be  included in the action in order to be invariant under the T-duality. The two multiplets with unknown coefficients are unphysical and do not contribute to S-matrix.  

The actions \reef{DBI2} and \reef{DBI3} are invariant under S-duality only at the on-shell level. If one would like to find an action which is invariant under off-shell   S-duality, one should first impose S-duality and then T-duality. We have done this calculation and found the following result:
\beqa
S&=&-\frac{\pi^2\alpha'^2T_p}{48}\int d^{p+1}\sigma\sqrt{-\det(\tG_{ab}+F_{ab} )}\bigg[-Q^a {}_a {}^b {}^c {}^d {}^e Q_c {}_b {}^f {}_f {}_d {}_e + 
 \frac{1}{2} Q^a {}^b {}^c {}_b {}^d {}^e Q_c {}_d {}^
   f {}_e {}_a {}_f \nonumber\\&&- 
 \frac{3}{2} Q^a {}^b {}^c {}_b {}^d {}^e Q_d {}_c {}^f {}_f {}_a {}_e   + 
 8 Q^a {}^b {}^c {}_b {}_c {}^d Q_d {}^e {}^f {}_e {}_a {}_f + 
\frac{7}{2} Q^a {}_a {}^b {}^c {}_b {}_c Q^d {}_d {}^e {}^
   f {}_e {}_f + 
 2 Q^a {}_a {}^b {}^c {}_b {}^d Q^e {}_c {}^f {}_f {}_d {}_e \nonumber\\&&+ 
\frac{1}{4} Q^a {}^b {}^c {}_a {}^d {}^e Q^
   f {}_b {}_d {}_f {}_c {}_e - 
 2 Q^c {}_c {}^d {}^
   e {}_d {}_e \Omega_a {}_b {}_i \Omega^a {}^
   b {}^i + 
 2 Q^d {}_b {}^e {}_d {}_c {}_e \Omega_a {}^
   c {}_i \Omega^a {}^b {}^i - 
 4 Q^d {}_d {}^e {}_b {}_c {}_e \Omega_a {}^
   c {}_i \Omega^a {}^b {}^i \nonumber\\&&- 
 2 Q^c {}_c {}^d {}^e {}_d {}_e \Omega^a {}_a {}^
   i \Omega^b {}_b {}_i - 
 2 Q^d {}_b {}^e {}_d {}_c {}_e \Omega^a {}_a {}^
   i \Omega^b {}^c {}_i - 
 4 Q^d {}_d {}^e {}_b {}_c {}_e \Omega^a {}_a {}^
   i \Omega^b {}^c {}_i + 
 4 Q_a {}_b {}^e {}_c {}_d {}_e \Omega^a {}^b {}^
   i \Omega^c {}^d {}_i\nonumber\\&& + 
 4 Q^e {}_a {}_e {}_c {}_b {}_d \Omega^a {}^b {}^
   i \Omega^c {}^d {}_i\bigg]\labell{ST}
\eeqa
where $Q$ is given in \reef{Q}. It is obviously invariant under linear S-duality for D$_3$-brane. The Levi-Civita tensors in above action should be evaluated for D$_3$-brane to find the corresponding contractions between $\Omega$ and $\prt F$. Then the result should be extended to arbitrary D$_p$-brane. We have found that the above couplings are consistent with the 4-point S-matrix element. In this case   there is one unphysical multiplet with unknown coefficient. The above action and the couplings in \reef{DBI2}, \reef{DBI3} looks very different. However, since both of them satisfy the same S-matrix element, they are related by field redefinitions. 

The action \reef{DBI1} includes  all infinite number of scalar couplings through the expansion of the inverse of the pull-back metric in the action.   So one may expect that the  consistency of the couplings with the dualities  and with \reef{DBI1} can be extended to all order of gauge fields. The S-duality is non-linear, and there are indications that D$_3$-brane effective action is   invariant under nonlinear S-duality only at the equations of motion level \cite{Gibbons:1995ap,Chemissany:2006qd},  so we do not expect the effective action at higher orders to be consistent with the nonlinear S-duality. In general, we do not expect them    to be   invariant under   linear S-duality even at the on-shell level. To see this, consider the   $\alpha'$ expansion of an S-matrix element which can be separated into two parts. One part includes massless poles and the other part includes contact terms, \ie
\beqa
A&=&A_{\rm pole}+A_{\rm contact}
\eeqa
 The linear S-duality transformation which is on the gauge field strength, \ie $F\rightarrow *F$, may transform  $A_{\rm pole}$ to $A_{\rm contact}$, so the contact terms which   produce the effective action,  may   not satisfy  the S-dual Ward identity\footnote{  It has been observed in \cite{Babaei-Aghbolagh:2013hia} that the combination of massless poles and contact terms of the S-matrix element of six gauge fields at the leading order of $\alpha'$ which is reproduced by the DBI action, is invariant under linear S-duality.}.   This is not the case for four gauge fields that we have considered in \reef{ST} because there is no massless pole for 4-point function. The S-matrix elements of one massless closed string and two NS states also have no massless poles, so one can  use the consistency of the couplings with the linear S-duality and T-duality  to find the physical couplings \cite{Jalali:2015xca}. 

On the other hand, the T-duality transformation which is on gauge field, \ie $A_y\rightarrow \chi^y$, does not transform $A_{\rm pole}$ to $A_{\rm contact}$, so the contact terms   always satisfy  the T-dual Ward identity. As a result, one expects the consistency of the higher order couplings with T-duality to be a valid constraint at higher order fields.  Therefore, we expect   the  consistency of the couplings with the T-duality  and with \reef{DBI1} to be extended to all order of gauge fields.

The T-duality   may  fix the presence of $F_{ab}$  in the pull-back of the flat metric in \reef{DBI1} by extending it to the following expression:
\beqa
\tG_{ab}&\longrightarrow&  {G}_{ab}=\eta_{ab}+\prt_a\chi^i\prt_b\chi^j\eta_{ij}-F_{ac}F_{db}\tG^{cd}\labell{ext}
\eeqa
The above extension then produces the following extensions for the metrics $\tG^{ab}$ and $\widetilde{\bot}_{ij}$ which appear at various places in \reef{DBI1}:
\beqa
\tG^{ab}&\longrightarrow&  {G}^{ab}\labell{trans}\\
\widetilde{\bot}_{ij}&\longrightarrow& {\bot}_{ij}=\eta_{ij}-\eta_{ik}\eta_{jl}\prt_a\chi^k\prt_b\chi^l\hat{G}^{ab}\nonumber
\eeqa
To verify that the replacement \reef{ext} is consistent with T-duality, suppose the D-brane is along the circle on which the T-duality is imposed. One can easily verify that the world-volume indices of $ {G}^{ab}$ in \reef{DBI1} are contracted with  the derivatives in $\Omega_{ab}{}^i$ or in $\prt_a\chi^k\prt_b\chi^l$, so they can not be  the Killing index $y$. On the other hand, when $a,b$ are not the Killing index, \ie $a=\ta,b=\tb$, using the prescription given in \cite{Garousi:2009dj}, one can   verify that  $ {G}^{\ta\tb}$ is invariant under T-duality. To the second order of fields, it is
\beqa
 {G}^{\ta\tb}&=&\eta^{\ta\tb}-\prt^{\ta}\chi^i\prt^{\tb}\chi^j\eta_{ij}+F^{\ta\tc}F ^{\td\tb}\eta_{\tc\td}+\prt^{\ta}A_y\prt^{\tb}A_y\eta^{yy}\nonumber\\
&\stackrel{T}{\longrightarrow}&\eta^{ab}-\prt^a\chi^{\ti}\prt^b\chi^{\tj}\eta_{\ti\tj}+F^{ac}F^{db}\eta_{cd}-\prt^a\chi^y\prt^b\chi^y\eta_{yy}\nonumber\\
&=&\eta^{ab}-\prt^a\chi^i\prt^b\chi^j\eta_{ij}+F^{ac}F^{db}\eta_{cd}\,=\, {G}^{ab}
\eeqa
where in the second  line we have used the T-duality transformation $A_y\rightarrow \chi^y$.   We have checked to the tenth order of fields and found that $ {G}^{\ta\tb}$  is invariant.   Therefore, the  replacement \reef{trans} forced by T-duality,   extends the couplings  \reef{DBI1} to include the constant gauge field strength to all orders. The resulting action, however, does not include the structure in which the two indices of the gauge field strength  contract with the second fundamental form, \ie the four world volume indices of $\Omega$'s contract with $F_{ab}F_{cd}$ or the eight world volume indices of $\Omega$'s contract with $F_{ab}F_{cd}F_{ef}F_{gh}$. Moreover, the world volume indices of $\Omega^i\Omega^j$'s may contract with $\prt_a\chi_i\prt_b\chi_j$. Since the world volume indices of $\Omega$'s are derivative indices, such couplings are invariant under T-duality only when the derivatives of the gauge field strength are zero.    The couplings  when the gauge field strength is not constant, may be found by requiring  the invariance of 6-field,    8-field, and higher couplings under T-duality. It would be interesting to use  these   conditions to find an action for  D$_p$-brane  which includes all orders of gauge fields and the transverse scalar fields. Such  action for D$_9$-brane has been  obtained in \cite{Wyllard:2000qe} via
string $\sigma$-model loop calculations using the boundary state operator language.    

 In the presence of the massless closed string fields, the S-duality is modified to $SL(2,R)$ symmetry. Unlike the transformation for the gauge field which is via its field strength, all other transformations involve only field potential. So   linear $SL(2,R)$ transformations for the massless closed string fields do not transform the massless pole, $A_{\rm pole}$, to the contact terms, $A_{\rm contact}$, of the S-matrix element. Therefore, we expect the effective actions which are produced by the contact terms of the S-matrix elements,  to be invariant under the linear $SL(2,R)$ transformations. There is a  subtlety, however,  for the S-matrix elements involving B-field. The B-field in the D-brane effective action appears in two ways. Either through its field strength $H$ or through the replacement $F\rightarrow F+B$. The massless poles in the S-matrix which are produced by the gauge fields, should be combined with some of the contact terms which are produced by $F+B$ terms,  to be able to rewrite the S-matrix element in terms of $H$ \cite{Garousi:2010bm,Garousi:2011ut}. Then the $H$-contact terms of the S-matrix element are the $H$-couplings in the effective action. These contact terms transforms to the RR $dC_{(2)}$-couplings under the $SL(2,R)$ transformation. On the other hand, the B-field in the effective action which appears as $F+B$, are expected to be invariant under the $S(L,R)$ transformation via the equations of motion  as in the DBI case \cite{Green:1996qg}. Apart from the $(F+B)$-couplings, we expect the D$_3$-brane effective action to be invariant under the $SL(2,R)$ transformation.

In general, the effective action of    D$_p$-brane is expected   to be invariant under   T-duality transformation, \eg the   DBI action is invariant under T-duality. The transformation is linear for the massless NS fields  whereas it is non-linear for   NSNS fields \cite{Bachas:1999um}. The  invariance of the curvature squared terms in O-plane action \cite{Bachas:1999um} under the non-linear T-duality, fixes all NSNS terms at order $\alpha'^2$   \cite{Robbins:2014ara,Garousi:2014oya}.    It would be  interesting to find all NSNS  and RR couplings in D-brane action at order $\alpha'^2$  by requiring their curvature squared terms  to be consistent with the nonlinear T-duality and with the $SL(2,R)$ transformations.

Finally, let us mention a subtlety in evaluating four Levi-Civita tensors in terms of metric that we have encountered  when we have been founding the couplings in \reef{ST}. Using \reef{Q} to write the couplings  in \reef{ST} in terms of $\prt F$, one finds terms that have four Levi-Civita tensors. They are produced by $QQ$ terms. For example, one encounters with the following term:
\beqa
\frac{1}{16} \psi_a {} _l {} _n \psi^
  a {} _g {} _h \psi_d {} _m {} _i \psi_e {} _j {} _k \epsilon_b {}^f {}^m {}^
  i \epsilon^b {}^c {}^g {}^h \epsilon_c {} _f {}^j {}^k \epsilon^d {}^e {}^l {}^n
	\eeqa
where $\psi_{abc}\equiv\prt_a F_{bc}$. To rewrite this term in terms of various contractions of $\psi$'s, one has to use the standard identity  $
\epsilon^{a_1a_2a_3a_4}\epsilon_{b_1b_2b_3b_4}=-\delta^{a_1}_{[b_1}\delta^{a_2}_{b_2}\delta^{a_3}_{b_3}\delta^{a_4}_{b_4]}$. However, there are three ways to pairs the Levi-Civita tensors. One of them is the following:
 \beqa
&&\frac{1}{16} \psi_a {} _l {} _n \psi^
  a {} _g {} _h \psi_d {} _m {} _i \psi_e {} _j {} _k (\epsilon_b {}^f {}^m {}^
  i \epsilon^b {}^c {}^g {}^h )(\epsilon_c {} _f {}^j {}^k \epsilon^d {}^e {}^l {}^n)=\frac{1}{2} \psi^a {} _a {}^f \psi_m {}^j {}^k \psi^m {} _f {}^i \psi_i {} _j {} _k + 
 \psi^a {} _a {}^f \psi_m {}^j {}^k \psi^m {} _f {}^i \psi_j {} _i {} _k \nonumber\\&&- 
 \psi_a {}^i {}^j \psi^a {}^f {}^m \psi_f {} _i {}^k \psi_k {} _m {} _j - 
\frac{1}{2} \psi_a {}^i {}^j \psi^a {}^f {}^m \psi_f {} _m {}^k \psi_k {} _i {} _j  - 
 \psi_a {} _f {}^i \psi^a {}^f {}^m \psi^j {} _m {}^k \psi_k {} _i {} _j + 
 \psi^a {} _a {}^f \psi_m {} _i {}^j \psi^m {} _f {}^i \psi^k {} _j {} _k\nonumber
\eeqa
where we have used the standard identity in the Levi-Civita tensors in each parenthesis. Another paring gives the following result:
 \beqa
&&\frac{1}{16} \psi_a {} _l {} _n \psi^
  a {} _g {} _h \psi_d {} _m {} _i \psi_e {} _j {} _k (\epsilon^d {}^e {}^l {}^
  n \epsilon^g {}^h {}^b {}^c )(\epsilon^m {}^i {} _b {}^f \epsilon^j {}^k {} _c {} _f)=\psi_a {}^m {}^i \psi^a {}^d {}^e \psi_d {} _m {}^l \psi_e {} _i {} _l - 
 \frac{1}{2} \psi^a {} _a {}^d \psi^e {} _d {} _e \psi_m {} _i {} _l \psi^m {}^i {}^
   l \nonumber\\&&+ 2 \psi_a {} _d {}^m \psi^a {}^d {}^e \psi_e {}^i {}^
   l \psi_i {} _m {} _l - 
\frac{1}{2}\psi_a {} _d {} _e \psi^a {}^d {}^e \psi^m {}^i {}^
   l \psi_i {} _m {} _l + 
 2 \psi^a {} _a {}^d \psi^e {} _d {}^m \psi_i {} _m {} _l \psi^i {} _e {}^l
\eeqa 
The two expressions are not identical! Even if one uses the Bianchi identity, the two results do not convert into each other!  In finding the couplings in \reef{ST} we have used the specific paring that the two Levi-Civita tensors in each $Q$ contract with each other.

{\bf Acknowledgments}:   This work is supported by Ferdowsi University of Mashhad under grant 2/33262(1393/12/05).

{\LARGE \bf Apendix}

In this appendix we review the low energy expansion of the S-matrix element of four gauge bosons in the superstring theory. The S-matrix element has been calculated  in \cite{Schwarz:1982jn}, 
\beqa
{\cal{A}}_{1234}&\sim&\frac{\Gamma(-s)\Gamma(-t)}{\Gamma(1+u)}K\labell{A}
\eeqa
where the Mandelstam variables are $s=- k_1\inn k_2$, $t=-  k_1\inn k_4$, $u=- k_1\inn k_3$ which satisfies $s+t+u=0$, and $K$ is the following  kinematic factor:
\beqa
K&=&- k_1\inn k_2(\z_1\inn k_4\z_3\inn k_2\z_2\inn\z_4+\z_2\inn k_3\z_4\inn k_1\z_1\inn\z_3+\z_1\inn k_3\z_4\inn k_2\z_2\inn\z_3+\z_2\inn k_4\z_3\inn k_1\z_1\inn\z_4)\nonumber\\
&&- k_2\inn k_3 k_2\inn k_4 \z_1\inn \z_2\z_3\inn\z_4+\{1,2,3,4\rightarrow 1,3,2,4\}+\{1,2,3,4\rightarrow 1,4,3,2\}\labell{kin}
\eeqa
which is   $stu$ symmetric.  In above amplitude $\alpha'=1/2$.
The $\alpha'$-expansion of the Gamma functions is 
\beqa
\frac{\Gamma(-s )\Gamma(-t )}{\Gamma(1+u)}&=&\frac{1}{st}-\frac{\pi^2}{6}-\z(3)(s+t)-\frac{\pi^4}{360}(4s^2+st+4t^2)+\cdots\nonumber
\eeqa
The total amplitude includes all non-cyclic permutation of the external states, \ie
\beqa
{\cal{A}}={\cal{A}}_{1234}+{\cal{A}}_{1243}+{\cal{A}}_{1324}+{\cal{A}}_{1342}+{\cal{A}}_{1423}+{\cal{A}}_{1432}
\eeqa
Using the relation $s+t+u=0$, one finds that  ${\cal{A}}$ has no   mssless pole. It becomes
 \beqa
{\cal A}&\sim& -\bigg[\pi^2+ \frac{\pi^4}{24}(t^2+s^2+u^2)+\cdots\bigg]K 
\eeqa
which produces only contact terms with   four, eight, and higher momenta. The four momenta terms are reproduced by the DBI action \reef{DBI} which  are proportional to $\pi^2$, and its eight momenta terms are reproduced by $\Omega^4$ terms \cite{Bachas:1999um} which are  proportional to $\pi^4$. The couplings for the transverse scalars can be found from the above couplings by using the condition that the scalar polarization is in transvers space, \ie $\z_i\inn k_j=0$.

\bibliographystyle{/Users/Nick/utphys} 
\bibliographystyle{utphys} \bibliography{hyperrefs-final}
\providecommand{\href}[2]{#2}\begingroup\raggedright
\endgroup

\end{document}